\documentclass[]{interact}

\pdfoutput=1
\usepackage[utf8]{inputenc}
\usepackage{amsmath, amssymb, graphicx,color}

\usepackage[numbers,sort&compress]{natbib}
\bibpunct[, ]{[}{]}{,}{n}{,}{,}

\definecolor{webblue}{rgb}{0, 0, 0.5} 
\usepackage[pdfpagelabels,  
plainpages=true,
bookmarks=true,
bookmarksnumbered=true,
hypertexnames=true,
breaklinks=true,%
linkcolor=webblue,]{hyperref}
\hypersetup{
colorlinks  = true,
linkcolor = webblue,
urlcolor = webblue,
citecolor = webblue,
pdfauthor   = {John F. Rudge},
pdfsubject = {},
pdftitle    = {Creep2},
pdfkeywords = {},
}

\renewcommand{\d}{\mathrm{d}}

\newcommand {\ddsq}[2]{\frac{\d^2 #1}{\d {#2}^2}}
\newcommand {\pdd}[2]{\frac{\partial #1}{\partial #2}}

\newcommand{\n}{\mathbf{n}}
\renewcommand{\x}{\mathbf{x}}

\newcommand{\dv}{\mathbf{d}}
\newcommand{\R}{\mathbf{R}}
\renewcommand{\v}{\mathbf{v}}
\newcommand{\om}{\boldsymbol{\omega}}

\newcommand{\uuline}[1]{\underline{\underline{#1}}}
\newcommand{\tensor}[1]{\uuline{\boldsymbol{#1}}}

\newcommand{\nablas}{\nabla_\perp}

\newcommand{\domega}{\Gamma^{\times}}
\newcommand{\dsigma}{\sigma^{\times}}

\begin{document}
\title{A micropolar continuum model of diffusion creep}
\author{John F. Rudge}
\author{
\name{John F. Rudge\thanks{Email: jfr23@cam.ac.uk}}
\affil{Bullard Laboratories, Department of Earth Sciences, Madingley Road, Cambridge, CB3 0EZ, UK.}
}

\maketitle

\begin{abstract}

Solid polycrystalline materials undergoing diffusion creep are usually described by Cauchy continuum models with a Newtonian viscous rheology dependent on the grain size. Such a continuum lacks the rotational degrees of freedom needed to describe grain rotation. Here we provide a more general continuum description of diffusion creep that includes grain rotation, and identifies the deformation of the material with that of a micropolar (Cosserat) fluid. We derive expressions for the micropolar constitutive tensors by a homogenisation of the physics describing a discrete collection of rigid grains, demanding an equivalent dissipation between the discrete and continuum descriptions. General constitutive laws are derived for both Coble (grain-boundary diffusion) and Nabarro-Herring (volume diffusion) creep. Detailed calculations are performed for a two-dimensional tiling of irregular hexagonal grains, which illustrates a potential coupling between the rotational and translational degrees of freedom. If only the plating out or removal of material at grain boundaries is considered, the constitutive laws are degenerate: modes of deformation that involve pure tangential motion at the grain boundaries are not resisted. This degeneracy can be removed by including the resistance to grain-boundary sliding, or by imposing additional constraints on the deformation.  
\end{abstract}


\section{Introduction}

At high temperatures, solid polycrystalline materials can deform by diffusion creep, where defects within the crystalline lattice move by diffusion. At scales much larger than the grain scale the material behaves as if it were a Newtonian viscous fluid, with an effective shear viscosity which depends on the grain size \citep{Coble1963,Nabarro1948,Herring1950,Lifshitz1963}. At the microscale individual grains can be considered as rigid bodies, which interact by the plating out or removal of material at grain boundaries, leading to a macroscale strain. Rigid bodies have both translational (velocity) and rotational (angular velocity) degrees of freedom to describe their motion. However, when a material is treated as a Newtonian viscous fluid at the macroscale, the microscale rotational degrees of freedom are lost, as the classical Cauchy continuum is based on point particles with only translational degrees of freedom. 

The aim of this paper is to present a model of diffusion creep that goes beyond 
the classical Cauchy continuum model, and instead identifies an appropriate 
micropolar (Cosserat) continuum. A micropolar continuum contains both 
translational and rotational degrees of freedom, and allows one to better 
describe phenomena associated with grain rotations and anisotropic 
microstructure. Micropolar models are used in a diverse range of disciplines 
e.g. to describe masonry \citep{Godio2017,Trovalusci2003}, granular media 
\citep{Vardoulakis2019}, the motion of fault-bounded blocks 
\citep{Twiss1993,Twiss2009}, rotational seismology \citep{Abreu2017,Abreu2018}, 
disclinations \citep{Cordier2014}, but as yet do not appear to have been used to 
describe diffusion creep.

The present work is a natural generalisation of the work of Wheeler \citep{Wheeler2010}, who studied the two-dimensional anisotropic Cauchy continuum arising from a periodic array of irregular hexagonal grains undergoing Coble (grain-boundary diffusion) creep. As will be discussed in detail later, the Cauchy continuum of Wheeler  \citep{Wheeler2010} can be seen as a reduced version of the more general micropolar continuum considered here.  


The manuscript is organized as follows. Section \ref{sec:micropolar} gives a brief overview of the governing equations of a micropolar continuum. Section \ref{sec:kinematics} describes the microscale kinematics of a collection of rigid grains. Section \ref{sec:diffusion} describes the microscale physics of diffusion. Section 
\ref{sec:dissipation} determines the dissipation associated with diffusion creep, and uses this as a basis for identifying the constitutive tensors for the equivalent micropolar continuum. Section \ref{sec:dimensionless} rescales the equations for the constitutive tensors into a dimensionless form. Simplified equations for Coble (grain-boundary diffusion) creep are given in section
\ref{sec:simplecoble}, followed by specific calculations for hexagonal grains in section \ref{sec:coblehex}. An important degeneracy in the constitutive laws in discussed in section \ref{sec:degeneracy}, and final conclusions are in section \ref{sec:conclusions}. Three appendices give additional mathematical details on the resistance to grain-boundary sliding, and simplifications of the micropolar continuum under additional assumptions.

%
%
%
%
%
%
%
%
%

%
%

\section{Micropolar continuum}\label{sec:micropolar}

In this section we briefly review the governing equations for a micropolar 
continuum, and a more detailed description can be found in 
e.g. \citep{Eringen1966,Eremeyev2013,Stefanou2017,Vardoulakis2019,Lukaszewicz1999}. A micropolar (Cosserat) 
continuum represents a continuous collection of particles which have both 
translational (velocity field $\mathbf{v}$) and rotational (microrotation-rate
field $\boldsymbol{\omega})$ degrees of freedom. At each point in the continuum 
we define both a force-stress tensor $\tensor{\sigma}$ and a couple-stress 
tensor $\tensor{\chi}$. In the absence of body-forces and body-couples, and 
with negligible inertia, the governing equations of a micropolar continuum are
\begin{gather}
 \sigma_{ij,j}  = 0,  \label{eq:micropolar_mom} \\
 \chi_{ij,j} - \epsilon_{ijk} \sigma_{jk}  = 0, \label{eq:micropolar_ang_mom}
\end{gather}
where $(\cdot)_{,j} \equiv \partial (\cdot) / \partial x_j$ and $\epsilon_{ijk}$ is the Levi-Civita symbol. Equation 
\eqref{eq:micropolar_mom} is the balance of linear momentum, and is identical to 
that of the classical Cauchy continuum. For a micropolar continuum the 
force-stress tensor $\tensor{\sigma}$ need not be symmetric, and one must be 
careful in how one assigns the indices. Here we adopt the convention that the 
traction vector $t_i$ on an infinitesimal patch with surface area $\d S$ 
and normal $n_j$ is given by $t_i = \sigma_{ij} n_j$. Similarly the moment 
vector $m_i$ is given by $m_i = \chi_{ij} n_j$. Equation 
\eqref{eq:micropolar_ang_mom} is the balance of angular momentum: in the absence 
of couple-stresses this reduces to usual statement for the Cauchy continuum that 
the force-stress tensor is symmetric. But here we will not assume vanishing 
couple-stresses. 

The balance of linear and angular momentum can equivalently be written as 
integral statements. The linear momentum statement is the force-balance
\begin{gather}
 \int \sigma_{ij} n_j \; \d S = 0,
\end{gather}
and the angular momentum statement is the torque-balance
\begin{gather}
 \int \chi_{ij} n_j + \epsilon_{ijk} x_j \sigma_{kl} n_l \; \d S  = 0, 
\end{gather}
where the integrals are over the bounding surface $S$ of any arbitrary volume 
$V$ of the continuum, and $\mathbf{x}$ represents the 
position vector. By dotting equation \eqref{eq:micropolar_mom} with 
$\mathbf{v}$ and 
\eqref{eq:micropolar_ang_mom} with $\boldsymbol{\omega}$ and integrating over a 
volume $V$, the following equation of mechanical energy balance can be 
obtained 
\begin{gather}
 \int v_i \sigma_{ij} n_j + \omega_{i} \chi_{ij} n_j \; \d S = \int \Gamma_{ij} 
\sigma_{ij} + K_{ij} \chi_{ij} \; \d V. \label{eq:mech_energy}
\end{gather}
The left-hand side in \eqref{eq:mech_energy} represents the rate of working by 
tractions and moments at 
the boundary of the given volume, and the right-hand side represents the 
internal power. The tensors $\tensor{\Gamma}$ and $\tensor{K}$ are the 
natural measures of deformation rate in a micropolar continuum, and are defined by
\begin{gather}
 \Gamma_{ij} = v_{i,j} + \epsilon_{kij} \omega_k, \label{eq:stain_def} \\
 K_{ij} = \omega_{i,j}. \label{eq:K_def}
 \end{gather}
These tensors can be shown to be objective measures of deformation rate. Note 
that unlike the traditional strain-rate tensor, the Cosserat tensor 
$\tensor{\Gamma}$ is not symmetric in general. However, it can be decomposed 
into symmetric and antisymmetric parts as
\begin{gather}
 \Gamma_{ij} = \Gamma_{(ij)} + \Gamma_{[ij]}, \label{eq:gam_as1}\\
 \Gamma_{(ij)} = \frac{1}{2} \left( v_{i,j} + v_{j,i} \right), \label{eq:gam_as2} \\
 \Gamma_{[ ij]} = \epsilon_{ijk} \left( \omega_k - \tfrac{1}{2} 
\Omega_k \right), \label{eq:rel_micro}
\end{gather}
where $\boldsymbol{\Omega}$ represent the classical vorticity
\begin{gather}
 \Omega_k = (\nabla \times \mathbf{v})_k = \epsilon_{klm}{v_{m, l}}.
\end{gather}
Round brackets $(ij)$ are used to represent the symmetric part of a 
tensor, and square brackets $[ij]$ are used to represent the 
antisymmetric part. The additional measures of deformation-rate a micropolar 
continuum has over a classical Cauchy continuum are the relative 
microrotation-rate measure in \eqref{eq:rel_micro} and the gradient in 
microrotation-rate in \eqref{eq:K_def}. 

Finally, constitutive laws are needed to relate the force- and couple-stress 
tensors to the measures of deformation rate. The most general linear constitutive 
laws take the form
\begin{gather}
 \sigma_{ij} = C_{ijkl} \Gamma_{kl} + B_{ijkl} K_{kl}, \label{eq:con1} \\
  \chi_{ij} = B_{klij} \Gamma_{kl} + D_{ijkl} K_{kl}, \label{eq:con2}
\end{gather}
with fourth-rank tensors $C_{ijkl}$, $B_{ijkl}$ and $D_{ijkl}$ describing the 
resistance of the medium to deformation. These tensors satisfy the major 
symmetries $C_{ijkl} = C_{klij}$ and  $D_{ijkl} = D_{klij}$. $B_{ijkl}$ is formally a pseudo-tensor, whereas $C_{ijkl}$ and $D_{ijkl}$ are proper tensors. With the 
constitutive laws in \eqref{eq:con1} and \eqref{eq:con2} the viscous 
dissipation per unit volume in \eqref{eq:mech_energy} can be written as
\begin{align}
 \Psi &\equiv \sigma_{ij} \Gamma_{ij} + \chi_{ij} K_{ij} \nonumber \\
 &= \Gamma_{ij} C_{ijkl} \Gamma_{kl} + 2 \Gamma_{ij} B_{ijkl} K_{kl} + 
K_{ij} D_{ijkl} K_{kl}. \label{eq:mpolar_diss}
\end{align}

The main aim 
of the present work is to determine expressions for the tensors $C_{ijkl}$, 
$B_{ijkl}$ and $D_{ijkl}$ from a homogenisation of the grain-scale physics 
describing diffusion creep.

\section{Kinematics of rigid grains}\label{sec:kinematics}

Our microscale model consists of a set of identical rigid grains in motion, where the 
relative motion normal to grain boundaries is accommodated by the plating out or removal of material. Each 
grain has a reference point $\x_i$, which we will choose to be the centroid 
of the grain. The rigid body motion of the grain is described by the velocity 
$\v_i$ of the reference point, and the angular velocity $\om_i$ about the 
reference point $\x_i$. The velocity $\v$ inside each grain is given by 
\begin{equation}
 \v = \v_i + \om_i \times \left( \x - \x_i \right)
\end{equation}
where $\x$ is the position vector.

\begin{figure}
\centering
 \includegraphics[width=0.5\columnwidth]{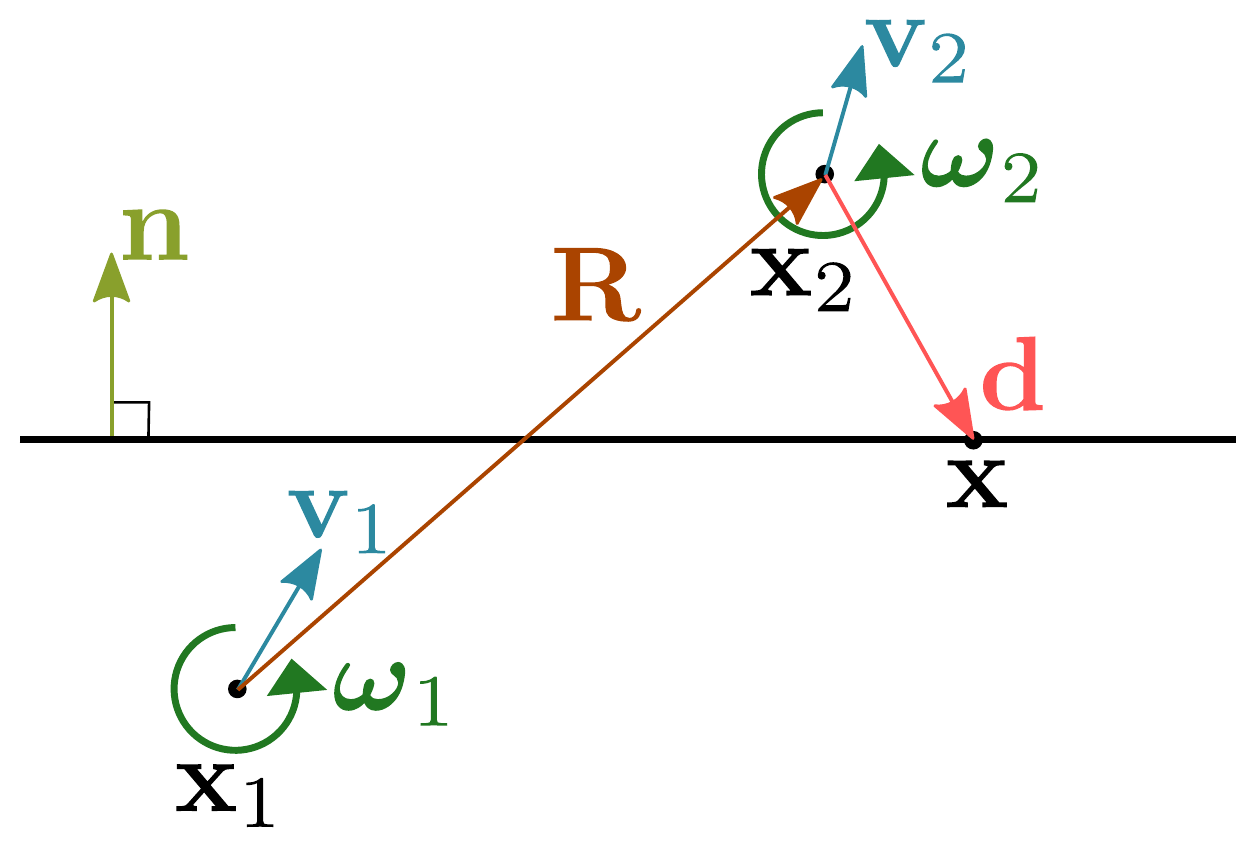}
 \caption{The geometry of a grain boundary.}
 \label{fig:grain_boundary}
\end{figure}

Consider the grain boundary between two grains, which we label as 1 and 2 (\autoref{fig:grain_boundary}). The 
difference in velocities between the two grains can be written as
\begin{equation}
 \Delta \v = \left(\v_2 - \v_1\right) - \om_1 \times \left(\x_2 - \x_1 \right) 
+ \left(\om_2 - \om_1 \right) \times \left(\x - \x_2 \right). \label{eq:deltav}
\end{equation}
Let $\R = \x_2 - \x_1$ be the vector joining the two grain centres. Suppose 
that the difference in velocities of the centroids and angular velocities is 
linearly related to $\R$, such that 
\begin{gather}
 \v_2 - \v_1 = \tensor{D} \cdot \R, \\
 \om_2 - \om_1 = \tensor{K} \cdot \R,
\end{gather}
where $\tensor{D}$ is a second rank tensor and $\tensor{K}$ is a 
second rank 
pseudo-tensor. After homogenisation we will identify $\tensor{D}$ with the 
velocity gradient tensor, and $\tensor{K}$ with the gradient of 
microrotation-rate.
Equation \eqref{eq:deltav} can then be written as
\begin{equation}
 \Delta \v = \tensor{D} \cdot \R  - \om_1 \times \R 
+ \left(\tensor{K} \cdot \R \right) \times \left(\x - \x_2 \right). 
\label{eq:deltav2}
\end{equation}
It is helpful to rewrite \eqref{eq:deltav2} in terms of the natural 
deformation-rate 
measure $\tensor{\Gamma}$, defined by
\begin{equation}
 \Gamma_{ij} = D_{ij} + \epsilon_{ijk} \omega^1_k,
\end{equation}
in a similar manner to the definition for the micropolar continuum in 
\eqref{eq:stain_def}. 
Equation \eqref{eq:deltav2} then becomes
\begin{equation}
 \Delta \v = \tensor{\Gamma} \cdot \R 
+ \left(\tensor{K} \cdot \R \right) \times \dv \label{eq:deltav3}
\end{equation}
where $\dv = \x - \x_2$. In diffusion creep, relative motion normal to the 
grain boundary is accommodated by plating. What is most of interest is then the 
normal component of \eqref{eq:deltav3}, given by
\begin{align}
 \Delta \v \cdot \n &= \n \cdot \tensor{\Gamma} \cdot \R 
+ \left(\dv \times \n \right) \cdot \left(\tensor{K} \cdot \R \right) \\
&= \tensor{\Gamma} : \R \n + \tensor{K}: \R \left(\dv \times \n \right)
\end{align}
We will use the symbol $\dot{r}$ to denote the plating rate on a given grain 
boundary for a given grain. If we assume that material is plated out 
symmetrically,
\begin{align}
 \dot{r} = \frac{1}{2} \Delta \v \cdot \n =\frac{1}{2} \left( \tensor{\Gamma} : 
\R \n +\tensor{K}: \R 
\left(\dv \times \n \right) \right) \label{eq:plating}
\end{align}
The above equation represents the key result of the kinematics: it gives 
the linear relation between the plating rate and the deformation-rate measures 
$\tensor{\Gamma}$ and $\tensor{K}$.

\section{Diffusion}\label{sec:diffusion}

Plating out or removal of material at grain boundaries can only occur by the 
diffusion of vacancies, either diffusion within the grains (leading to 
Nabarro-Herring creep), or diffusion along the grain boundaries 
(leading to Coble creep). In this section we will determine the relationships 
between the plating rates and the variations in vacancy concentration (or 
equivalently, chemical potential) that arise from the physics of diffusion. The 
motion of each grain relative to its neighbours is assumed to be well-described 
by constant values of the deformation rate measures $\tensor{\Gamma}$ and 
$\tensor{K}$ for each grain. Since the vacancy concentration is linearly 
dependent on the plating rate, and the plating rate depends linearly on the 
deformation rate measures 
$\tensor{\Gamma}$ and $\tensor{K}$, the vacancy concentration $c$ also depends 
linearly on $\tensor{\Gamma}$ and $\tensor{K}$. This linear relationship can be written as
\begin{equation}
 c = \gamma_{kl} \Gamma_{kl} + \nu_{kl} K_{kl} \label{eq:c}
\end{equation}
where $\gamma_{kl}$ and $\nu_{kl}$ are spatially-variable second-rank tensors that are determined by the 
geometry of the grains and the equations describing the diffusion.

\subsection{Nabarro-Herring creep (volume diffusion)}

In Nabarro-Herring creep diffusion only takes place within the body of the 
grains. The flux of vacancies $\mathbf{j}_\text{v}$ within each grain is described by Fick's law,
\begin{equation}
 \mathbf{j}_\text{v} = - D_\text{v} \nabla c, \label{eq:fluxdef}
\end{equation}
where $D_\text{v}$ is the diffusivity of vacancies, and $c$ is the concentration 
of vacancies. With a quasi-steady approximation, conservation of vacancies 
within the grain can be expressed as ${\nabla\cdot
\mathbf{j}_\text{v}=0}$, from which it follows that the concentration of vacancies $c$ satisfies Laplace's equation
\begin{equation}
 \nabla^2 c = 0  \text{ in }V, \label{eq:nh1}
\end{equation}
where $V$ is the volume of a grain.  Consideration of the boundary flux required 
to produce a certain plating rate yields
\begin{equation}
 \dot{r} = - \Omega \mathbf{j}_\text{v} \cdot \mathbf \n 
\label{eq:plating_flux}
\end{equation}
where $\Omega$ is the atomic volume, and $\mathbf{n}$ is the outward normal to the grain. Substitution of \eqref{eq:fluxdef} into \eqref{eq:plating_flux} yields the boundary condition
\begin{equation}
 \pdd{c}{n} = \frac{\dot{r}}{\Omega D_\text{v}} \text{ on } S \label{eq:nh2}
\end{equation}
where $S$ is the surface of the grain. For a given plating rate distribution 
$\dot{r}$, the linear equations \eqref{eq:nh1} and \eqref{eq:nh2} can be solved 
to give the concentration of vacancies $c$ inside the grain. The plating rate in 
turn depends linearly on the deformation rate measures $\tensor{\Gamma}$ and 
$\tensor{K}$ through the kinematic relationship of \eqref{eq:plating}, where 
$\tensor{\Gamma}$ and $\tensor{K}$ are assumed to be constant for the grain. 
The overall relationship between the concentration $c$ and the deformation rate 
measures can be written as a general linear relationship of the form in 
\eqref{eq:c}. Combining \eqref{eq:plating}, \eqref{eq:c}, \eqref{eq:nh1}, and 
\eqref{eq:nh2} 
then yields the following problems for the tensors $\gamma_{kl}$ and $\nu_{kl}$,
\begin{gather}
 \nabla^2 \gamma_{kl} = 0 \text{ in }V, \\
\pdd{\gamma_{kl}}{n} = \frac{1}{2 \Omega D_\text{v}} n_k R_l
\text{ on } S,
\end{gather}
and
\begin{gather}
 \nabla^2 \nu_{kl} = 0 \text{ in }V, \\
\pdd{\nu_{kl}}{n} = \frac{1}{2 \Omega D_\text{v}} \left(\dv \times \n \right)_k R_l
\text{ on } S.
\end{gather}

\subsection{Coble creep (grain boundary diffusion)}

Diffusion along grain boundaries can be described by Fick's law in the form
\begin{equation}
 \mathbf{j}_\text{v} = - D^\text{gb}_\text{v} \nablas c, 
\label{eq:fluxdef_c}
\end{equation}
where $\mathbf{j}_\text{v}$ is the flux of vacancies, $D^\text{gb}_\text{v}$ is 
the diffusivity of vacancies along the grain boundary, and $c$ is the concentration of vacancies. $\nablas$ represents the surface gradient operator (the gradient operator with the component normal to the grain boundary removed). Conservation of mass can be written as  
\begin{equation}
\tfrac{1}{2} \Omega \delta \nablas \cdot \mathbf{j}_\text{v} = 
\dot{r} \text{ on } S 
\label{eq:coblemass}
\end{equation}
where $\delta$ is the grain boundary thickness. The factor of 1/2 arises because each grain boundary borders two
grains. Substitution of \eqref{eq:fluxdef_c} into \eqref{eq:coblemass} yield the governing equation
\begin{equation}
- \tfrac{1}{2}  \Omega \delta D^\text{gb}_\text{v} \nablas^2 c = 
\dot{r} \text{ on } S  \label{eq:coble_main}
\end{equation}
where $\nablas^2$ represents the surface Laplacian operator. Equation \eqref{eq:coble_main} has to be 
supplemented by boundary conditions at the junctions where grain boundaries 
meet. Combining \eqref{eq:coble_main} with \eqref{eq:plating} and \eqref{eq:c} leads 
to the following problems for the tensors $\gamma_{kl}$ and $\nu_{kl}$,
\begin{equation}
-  \nablas^2 \gamma_{kl} = \frac{1}{\Omega \delta D^\text{gb}_\text{v}} n_k R_l 
\text{ on } S ,
\end{equation}
and
\begin{equation}
-  \nablas^2 \nu_{kl} = \frac{1}{\Omega \delta D^\text{gb}_\text{v}} \left(\dv 
\times \n \right)_k R_l \text{ on } S.
\end{equation}

\section{Dissipation}\label{sec:dissipation}

The upscaling technique used in this work is based on the notion of a 
homogeneous equivalent continuum \citep{Charalambakis2010}. The discrete 
collection of grains is replaced with a micropolar continuum that (1) shares 
the 
same kinematics as the discrete collection and (2) shares the same dissipation 
of energy. The kinematic mapping is achieved by the continuum field for the 
velocities $\v$ and angular velocities $\om$ matching the grain velocities 
$\v_i$ and $\om_i$ at the reference points $\x_i$ of each grain. The energetics 
of diffusion can be described by the balance law
\begin{equation}
\Psi = \frac{k T}{c_0 \Omega} \frac{1}{V} \int \dot{r} c \; \d S 
\label{eq:orig_diss}
\end{equation}
where $\Psi$ is the dissipation and the right hand side represents the rate of 
working at the grain boundaries. $c_0$ is the equilibrium concentration of 
vacancies, $k$ is the Boltzmann constant, $T$ is temperature, and $V$ is 
the grain volume. The surface integral is over the boundary of the grain.  
Explicit integral expressions for the dissipation $\Psi$ due to Nabarro-Herring 
creep and Coble creep are given below in \eqref{eq:nh_diss} and 
\eqref{eq:coble_diss}.

The condition of equal dissipation between the discrete collection of grains 
and the continuum is imposed by demanding that the micropolar dissipation in 
\eqref{eq:mpolar_diss} matches that in the balance law \eqref{eq:orig_diss}. By 
substituting the linear relationships \eqref{eq:plating} and \eqref{eq:c} into 
the rate of working integral in \eqref{eq:orig_diss}, one can obtain 
expressions for the 
tensors $C_{ijkl}$, $B_{ijkl}$, and $D_{ijkl}$ as
\begin{align}
 C_{ijkl} &= \frac{k T}{2 c_0 \Omega V}  \int  n_i 
R_j  
 \gamma_{kl} \; \d S,  \label{eq:t1} \\
B_{ijkl} &= \frac{k T}{2 c_0 \Omega V}  \int n_i R_j \nu_{kl} \; \d 
S = \frac{k T}{2 c_0 \Omega V} \int \gamma_{ij} \left(\dv 
\times \n \right)_k R_l \; 
\d 
S,  \label{eq:t2} \\
D_{ijkl} &= \frac{k T}{2 c_0 \Omega V} \int \left(\dv 
\times \n \right)_i R_j \nu_{kl} \; \d S. \label{eq:t3}
\end{align}
Since the expression for the rate of working is the same for both 
Nabarro-Herring and Coble creep, the above expressions for the constitutive 
tensors are valid for both forms of creep. However, one can also express these 
tensors in alternative, equivalent forms based on the explicit integral 
expressions of the dissipation for the two forms of creep.

\subsection{Nabarro-Herring creep}

For Nabarro-Herring creep we can write the 
dissipation in \eqref{eq:orig_diss} explicitly as
\begin{equation}
\Psi \equiv \frac{k T D_\text{v}}{c_0 } \frac{1}{V} \int | \nabla c |^2 \; \d 
V, 
\label{eq:nh_diss}
\end{equation}
where the balance law in \eqref{eq:orig_diss} is as a consequence of 
the divergence theorem and the governing equations \eqref{eq:nh1} and 
\eqref{eq:nh2}. Matching \eqref{eq:nh_diss} with \eqref{eq:mpolar_diss} 
gives an 
alternative expression for the tensors for Nabarro-Herring creep as
\begin{gather}
C_{ijkl} = \frac{k T D_\text{v}}{c_0 } \frac{1}{V} \int \nabla \gamma_{ij} \cdot 
\nabla \gamma_{kl} \; \d V, \\
B_{ijkl} = \frac{k T D_\text{v}}{c_0 } \frac{1}{V} \int \nabla \gamma_{ij} \cdot 
\nabla \nu_{kl} \; \d V, \\
D_{ijkl} = \frac{k T D_\text{v}}{c_0 } \frac{1}{V} \int \nabla \nu_{ij} \cdot 
\nabla \nu_{kl} \; \d V.
\end{gather}
In this representation the major symmetries $C_{ijkl} = C_{klij}$ and $D_{ijkl} 
= D_{klij}$ are clear.

\subsection{Coble creep}

For Coble creep we can write the dissipation in \eqref{eq:orig_diss} explicitly 
as
\begin{equation}
\Psi \equiv \frac{k T \delta D^\text{gb}_\text{v}}{c_0 } \frac{1}{2 V} \int | 
\nablas c |^2 \; \d S, \label{eq:coble_diss}
\end{equation}
where the balance law in \eqref{eq:orig_diss} is a consequence of the 
divergence theorem and the governing equation in \eqref{eq:coble_main}. Matching 
\eqref{eq:coble_diss} with \eqref{eq:mpolar_diss} gives an 
alternative expression for the tensors for Coble creep as
\begin{gather}
C_{ijkl} = \frac{k T \delta D^\text{gb}_\text{v}}{c_0 } \frac{1}{2V} \int 
\nablas \gamma_{ij} \cdot 
\nablas \gamma_{kl} \; \d S, \\
B_{ijkl} = \frac{k T \delta D^\text{gb}_\text{v}}{c_0 } \frac{1}{2V} \int 
\nablas \gamma_{ij} \cdot 
\nablas \nu_{kl} \; \d S, \\
D_{ijkl} = \frac{k T \delta D^\text{gb}_\text{v}}{c_0 } \frac{1}{2V} \int 
\nablas \nu_{ij} \cdot 
\nablas \nu_{kl} \; \d S,
\end{gather}
where again the major symmetries are clear.

\section{Dimensionless equations}\label{sec:dimensionless}

It is helpful to make the equations dimensionless to separate out the part of 
the behaviour that can be determined simply by scaling, and that which 
describes the geometrical effects of the microstructure. If $d$ is a typical 
measure of grain size, then a scaling for a typical viscosity is $\eta_0 = k T 
d^2 /(\Omega D)$ for Nabarro-Herring creep and $\eta_0 = k T d^3 
/(\Omega \delta D^\text{gb})$ for Coble creep, where $D = \Omega c_0 D_\text{v}$ and $D^\text{gb} = \Omega c_0 D^\text{gb}_\text{v}$ are the self-diffusion coefficients. Dimensionless versions of the 
fourth-rank tensors can be obtained by scaling $C_{ijkl}$ by $\eta_0$, 
$B_{ijkl}$ by $d \eta_0$ and $D_{ijkl}$ by $d^2 
\eta_0$. In dimensionless form, \eqref{eq:t1}, \eqref{eq:t2}, \eqref{eq:t3} 
become
\begin{align}
 C_{ijkl} &= \frac{1}{2 V}  \int  n_i 
R_j  
 \gamma_{kl} \; \d S,  \label{eq:t1_n} \\
B_{ijkl} &= \frac{1}{2 V}  \int n_i R_j \nu_{kl} \; \d 
S = \frac{1}{2 V} \int \gamma_{ij} \left(\dv 
\times \n \right)_k R_l \; 
\d 
S,  \label{eq:t2_n} \\
D_{ijkl} &= \frac{1}{2 V} \int \left(\dv 
\times \n \right)_i R_j \nu_{kl} \; \d S. \label{eq:t3_n}
\end{align}

\subsection{Nabarro-Herring creep (dimensionless)}
The corresponding dimensionless problem for Nabarro-Herring creep is
\begin{gather}
 \nabla^2 \gamma_{kl} = 0 \text{ in }V, \\
\pdd{\gamma_{kl}}{n} = \frac{1}{2} n_k R_l
\text{ on } S,
\end{gather}
and
\begin{gather}
 \nabla^2 \nu_{kl} = 0 \text{ in }V, \\
\pdd{\nu_{kl}}{n} = \frac{1}{2} \left(\dv \times \n \right)_k R_l
\text{ on } S.
\end{gather}


\subsection{Coble creep (dimensionless)}

The corresponding dimensionless problem for Coble creep is
\begin{equation}
-  \nablas^2 \gamma_{kl} = n_k R_l  \label{eq:Coble_g} 
\text{ on } S ,
\end{equation}
and
\begin{equation}
-  \nablas^2 \nu_{kl} = \left(\dv 
\times \n \right)_k R_l \text{ on } S. \label{eq:Coble_nu}
\end{equation}


\section{Simplified Coble creep calculations}\label{sec:simplecoble}

Further simplifications can be made for Coble creep, where with  
additional assumptions simpler expressions for the relevant tensors can be 
obtained. We will assume now that all grain boundaries are planar, and that the 
triple lines where different grain boundaries meet are at 
a constant chemical potential (``shorted'' in the language of 
Rudge \citep{Rudge2018a}). In the work of Rudge \citep{Rudge2018a}, shorted boundary conditions were introduced to mimic the effect of a small amount of melt lying along the triple lines. The melt acts as a fast path for diffusion, and allows for bulk deformation to occur. With shorted boundary conditions, the diffusion problem 
for each grain boundary is independent of the other boundaries. Let us label 
each grain boundary by an index $\alpha$. Since the normal vector $\n$ and the 
vector $\R$ joining centroids are constant over each grain boundary, the Coble 
creep problem in \eqref{eq:Coble_g} can be reduced to  
\begin{gather}
\gamma_{kl} = w \, n_k R_l, \\
-  \nablas^2 w= 1 \label{eq:dprob1}
\text{ on } S^{(\alpha)},
\end{gather}
where the shorted boundary conditions imply that $w$ vanishes along the 
bounding curve of the grain boundary. Similarly, the problem 
\eqref{eq:Coble_nu} can be reduced to 
\begin{gather}
\nu_{kl} = g_k R_l, \\
- \nablas^2 g_{k} = \left(\dv 
\times \n \right)_k  \text{ on } S^{(\alpha)}, \label{eq:dprob2}
\end{gather}
where the vector $g_{k}$ vanishes on the bounding curve of each grain boundary.

The fourth-rank tensors can then be expressed as
\begin{gather}
 C_{ijkl} = \frac{1}{2 V} \sum_{\alpha}  n_i^{(\alpha)}  \label{eq:C_red}
R_j^{(\alpha)} n_k^{(\alpha)} R_l^{(\alpha)} W^{(\alpha)},\\
 B_{ijkl} = \frac{1}{2 V} \sum_{\alpha}  n_i^{(\alpha)} 
R_j^{(\alpha)} F_k^{(\alpha)} R_l^{(\alpha)}, \label{eq:B_red}\\
 D_{ijkl} = \frac{1}{2 V} \sum_{\alpha}  G_{ik}^{(\alpha)}  R_j^{(\alpha)} 
R_l^{(\alpha)}, \label{eq:D_red}
\end{gather}
where the sums are over each of the grain boundaries for the given grain, and 
\begin{align}
 W^{(\alpha)} &= \int_{S^{(\alpha)}} w \; \d S, \label{eq:mom1} \\
F_k^{(\alpha)} &= \int_{S^{(\alpha)}} w \left(\dv \times \n \right)_k 
\; \d S = \int_{S^{(\alpha)}} g_k \; \d S, \label{eq:mom2} \\
 G_{ik}^{(\alpha)} &= \int_{S^{(\alpha)}} \left(\dv \times \n 
\right)_i g_k
\; \d S, \label{eq:mom3}
\end{align}
where it should be noted that the tensor $G_{ik}^{(\alpha)}$ is symmetric.

\section{Coble creep of hexagons}\label{sec:coblehex}

To give a concrete example, we now consider the specific case of Coble creep of 
a tiling of hexagonal grains. We restrict the motion to two-dimensions, so that 
$\mathbf{v} = (v_1, v_2, 0)$ and $\boldsymbol{\omega} = (0, 0, \omega_3)$. In 
two-dimensions $\Gamma_{3j} = \Gamma_{i3} = 0$, and the only non-zero 
components of $K_{ij}$ are $K_{31}$ and $K_{32}$. Thus only $C_{ijkl}$, 
$B_{ij3l}$, and $D_{3j3l}$ are non-zero for $i,j, k, l = 1,2$. With 
two-dimensional grains, the Coble creep diffusion problems are one-dimensional 
and are simple to solve analytically.  

\subsection{Regular hexagons}

The simplest hexagonal tiling is that of regular hexagons, with the wallpaper group \textit{p}6\textit{m}. With this hexagonal symmetry, Neumann's principle implies that the tensors $C_{ijkl}$ and $D_{3j3l}$ are isotropic, and the pseudo-tensor $B_{ij3l}$ vanishes. Moreover, for the regular hexagons, the vector $\R$ between 
centroids is in the same direction as the normal vector $\mathbf{n}$. It follows 
from \eqref{eq:C_red} that there are the further minor symmetries $C_{ijkl} = 
C_{jikl} = C_{ijlk}$, along with the Cauchy relation symmetry 
$C_{ijkl} = 
C_{ikjl}$. As a consequence of these symmetries, the 
constitutive tensors reduce to
\begin{gather}
 C_{ijkl} = \eta \left(\delta_{ij} \delta_{kl} + \delta_{ik} \delta_{jl} + 
\delta_{il} \delta_{jk} \right), \label{eq:C_reg_hex} \\
B_{ij3l} = 0, \label{eq:B_reg_hex} \\
D_{3j3l} = \mu \delta_{jl}, \label{eq:D_reg_hex} 
\end{gather}
in terms of two constants $\eta$ and $\mu$, where $\delta_{ij}$ represents the 
two-dimensional Kronecker delta. $\eta$ represents the effective 
shear viscosity. The effective bulk viscosity $\zeta = 2 \eta$, a consequence 
of the Cauchy relation symmetry \citep{Rudge2018a}. Solution of the diffusion problem 
leads to $\eta=1/144$ if lengths are made dimensionless on the 
perpendicular distance between opposite sides of the hexagon. This value is 
exact agreement with previous calculations (e.g. 
\citep{Spingarn1978,Cocks1990,Rudge2018a}). The new behaviour is captured by 
the constant $\mu$, which describes the resistance to relative rotation of the 
grains. As will be seen in the next section, $\mu = \eta/45$ in 
the dimensionless variables. 

\subsection{Irregular hexagons}\label{sec:hex}

\begin{figure}
\centering
 \includegraphics[width=0.7\columnwidth]{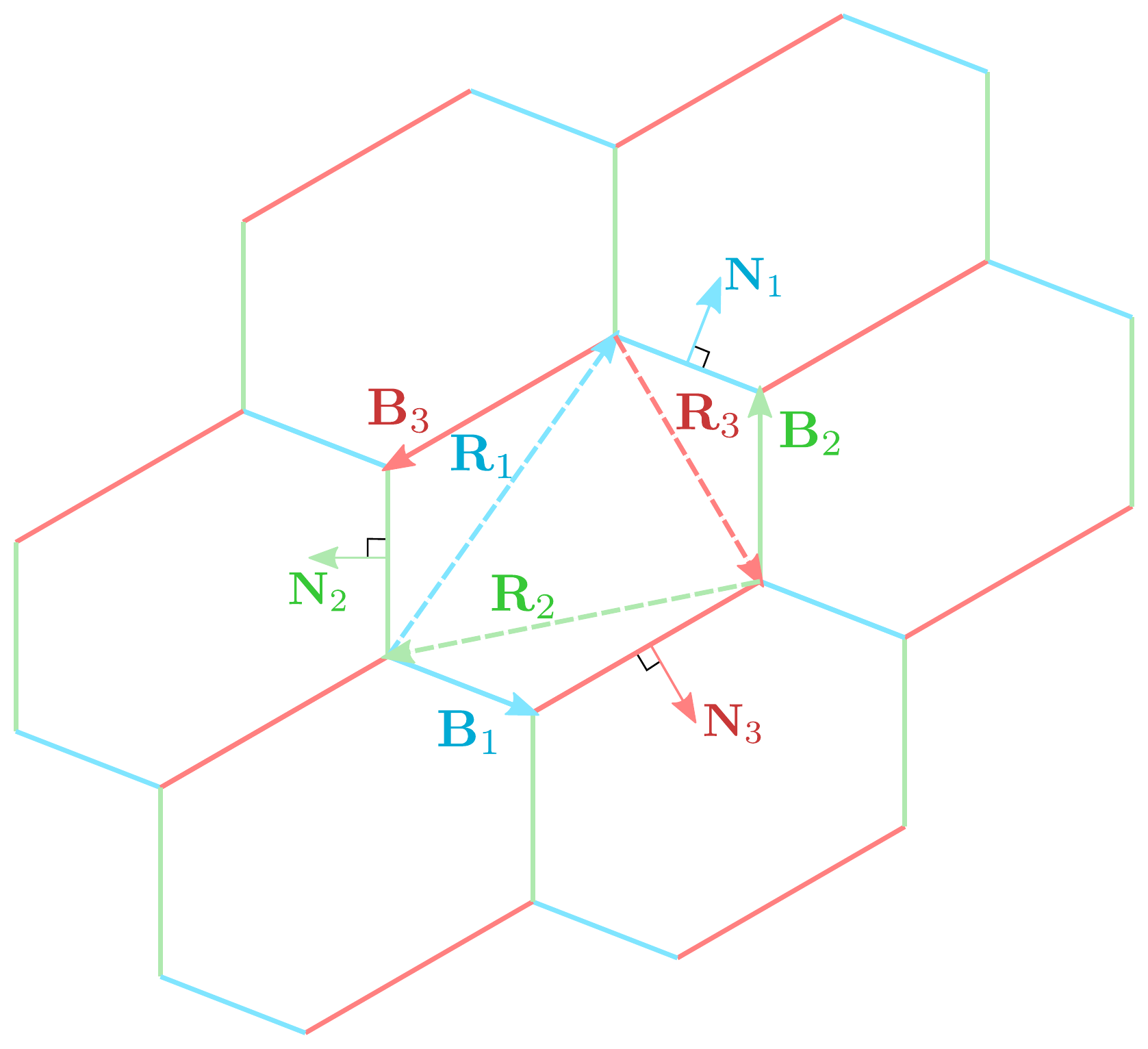}
 \caption{Periodic tiling of irregular hexagonal grains as considered by Wheeler \citep{Wheeler2010}. Note that the centroid of each grain, and the midpoint of each grain boundary is a rotation centre of order two (the tiling is invariant under a rotation by 180$^\circ$ about these points). The symmetry of the tiling is the wallpaper group \textit{p}2. All triple junctions in the tiling are indistinguishable owing to this symmetry.}
 \label{fig:wheeler_hex}
\end{figure}

Wheeler \citep{Wheeler2010} presented a thorough discussion of Coble creep for a periodic tiling 
of irregular hexagonal grains and determined the constitutive laws for an equivalent Cauchy continuum. In this section, we determine the corresponding
micropolar constitutive 
tensors for these irregular hexagons. The
grain 
shape is a hexagon with parallel sides, where the sides are given by the three 
vectors $\mathbf{B}_1$, $\mathbf{B}_2$, and $\mathbf{B}_3$ (\autoref{fig:wheeler_hex}). The 
vectors joining 
grain centroids are given by
\begin{gather}
 \mathbf{R}_1 = \mathbf{B}_2 - \mathbf{B}_3, \\ 
 \mathbf{R}_2 = \mathbf{B}_3 - \mathbf{B}_1, \\
 \mathbf{R}_3 = \mathbf{B}_1 - \mathbf{B}_2. 
\end{gather}
The area of the hexagon is
\begin{equation}
 V = \frac{1}{2} \sum_{i=1}^3 |  \mathbf{R}_i \times \mathbf{B}_i |,
\end{equation}
and unit normal vectors to the grain boundaries are given by
\begin{equation}
 \mathbf{N}_{f} = \hat{\mathbf{z}} \times \hat{\mathbf{B}}_f 
\end{equation}
for $f=1, 2, 3$, where  $\hat{\mathbf{B}}_f = \mathbf{B}_f / 
|\hat{\mathbf{B}}_f|$ is a unit vector.

\subsubsection{Diffusion problems}

Suppose the centroid of the grain is the origin of the coordinate system. The 
position vector along the edge with normal $\mathbf{N}_1$ in \autoref{fig:wheeler_hex} is
\begin{equation}
 \mathbf{x} = \frac{1}{2} \mathbf{R}_1 - s \hat{\mathbf{B}}_1, \quad 
-\frac{l_1}{2} \leq s \leq \frac{l_1}{2},
\end{equation}
where $l_1 = | \mathbf{B_1} |$. Now $\dv_1 = \x - \R_1$, from which it follows 
that
\begin{equation}
 \dv_1 \times \n =\left(\dv_1 \cdot \hat{\mathbf{B}}_1 \right)  \hat{\mathbf{z}} 
 = - \left(\frac{\mathbf{R}_1 \cdot \hat{\mathbf{B}}_1}{2} + s \right) 
\hat{\mathbf{z}}. 
\end{equation}
The diffusion problem in \eqref{eq:dprob1} becomes
\begin{equation}
 -\ddsq{w}{s} = 1, \quad w = 0 \text{ at } s= \pm \frac{l_1}{2}
\end{equation}
with solution
\begin{equation}
 w = \frac{1}{8} \left( l_1^2 - 4 s^2 \right). \label{eq:dsol1}
\end{equation}
The diffusion problem in \eqref{eq:dprob2} becomes
\begin{equation}
 -\ddsq{g_3}{s} = -\left(\frac{\mathbf{R}_1 \cdot \hat{\mathbf{B}}_1}{2} + s 
\right), \quad 
g_3 = 0 \text{ at } s= \pm \frac{l_1}{2}
\end{equation}
with solution
\begin{equation}
 g_3 = - \frac{1}{48}\left(3 \mathbf{R}_1 \cdot \hat{\mathbf{B}}_1 + 2 s 
\right) \left( l_1^2 - 4 s^2  \right). \label{eq:dsol2}
\end{equation}
From \eqref{eq:dsol1} and \eqref{eq:dsol2} it follows that the relevant moments 
in equations \eqref{eq:mom1}, \eqref{eq:mom2}, and \eqref{eq:mom3} are

\begin{gather}
W = \int_{-l_1/2}^{l_1/2} w \; \d s = \frac{l_1^3}{12}, \\
F_3 = \int_{-l_1/2}^{l_1/2} w \left(\dv_1 \times \n\right)_3 \; \d s =
 \int_{-l_1/2}^{l_1/2} g_3 \; \d s = - \frac{\left(\mathbf{R}_1 \cdot 
\mathbf{B}_1 \right)  l_1^2}{24}, \\
G_{33} = \int_{-l_1/2}^{l_1/2} g_3 \left(\dv_1 \times \n\right)_3 \; \d s = 
\frac{1}{720} \left( 15 \left(\mathbf{R}_1 \cdot \mathbf{B}_1 \right)^2  l_1 + 
l_1^5 \right).
\end{gather}
Thus from \eqref{eq:C_red}, \eqref{eq:B_red}, and \eqref{eq:D_red} the 
constitutive tensors are
\begin{gather}
 C_{ijkl} = \frac{1}{12 V} \sum_{f=1}^3 N^f_i R^f_j N^f_k R^f_l  l_f^3, 
\label{eq:C_hex} \\
B_{ij3l} = -\frac{1}{24 V} \sum_{f=1}^3 N^f_i R_j^f  R^f_l \left(\mathbf{R}_f 
\cdot 
\mathbf{B}_f \right) l_f^2,  \label{eq:B_hex}\\
D_{3j3l} = \frac{1}{720 V} \sum_{f=1}^3 R_j^f R_l^f  \left( 15 
\left(\mathbf{R}_f \cdot 
\mathbf{B}_f \right)^2  l_f + l_f^5 \right), \label{eq:D_hex}
\end{gather}
where $N_i^f$ denotes the $i^{\text{th}}$ component of the vector 
$\mathbf{N}_f$, and  $R_j^f$ denotes the $j^{\text{th}}$ component of the 
vector 
$\mathbf{R}_f$. 

For the case of regular hexagons, the expressions in \eqref{eq:C_hex}, 
\eqref{eq:B_hex}, and \eqref{eq:D_hex} simplify to those in 
\eqref{eq:C_reg_hex}, \eqref{eq:B_reg_hex}, and \eqref{eq:D_reg_hex}. For 
regular hexagons $\mathbf{R}_f \cdot 
\mathbf{B}_f = 0$. In dimensionless units where the perpendicular distance 
between sides is 1, $\mathbf{R}_f = \mathbf{N}_f$,  $V = \sqrt{3}/2$ and $l_f = 
1/ \sqrt{3}$. One can calculate $\eta$ in \eqref{eq:C_reg_hex} from the 
invariant $C_{iikk} = 1/18 = 8 \eta$, to yield $\eta = 1/144$. $\mu$ in 
\eqref{eq:D_reg_hex} can be calculated from the invariant $D_{3j3j}  = 1/ 3240 
= 
2 \mu$ and hence $\mu= 1/6480=\eta/45$. 

The constitutive tensors in \eqref{eq:C_hex}, \eqref{eq:B_hex}, and \eqref{eq:D_hex} can be seen as a generalisation of the constitutive tensors derived by Wheeler \citep{Wheeler2010}. Appendix \ref{sec:reduced} describes how the constitutive tensors here reduce to those of Wheeler \citep{Wheeler2010} with the additional assumptions that are made in that work.                                                                                                                            

\section{Degeneracy}\label{sec:degeneracy}

In this section we will show that the constitutive laws that have been derived  are in some sense unsatisfactory because they 
are degenerate. Ultimately this arises because there are modes of deformation 
that can occur without plating, and thus without any dissipation.

It is helpful to begin by noting the difference between the 
isotropic constitutive laws derived for the 2D tiling of regular hexagons in 
\eqref{eq:C_reg_hex}, \eqref{eq:B_reg_hex}, and \eqref{eq:D_reg_hex} and that 
of the most general two-dimensional linear isotropic micropolar continuum, 
which can be written in terms of four material constants as
\begin{gather}
 C_{ijkl} = \zeta \delta_{ij} \delta_{kl} +  \eta \left(
\delta_{ik} \delta_{jl} + 
\delta_{il} \delta_{jk} - \delta_{ij} \delta_{kl}  \right)  
+ \kappa \left(
\delta_{ik} \delta_{jl} - 
\delta_{il} \delta_{jk}   \right),  \label{eq:C_iso} 
\\
B_{ij3l} = 0, \label{eq:B_iso} \\
D_{3j3l} = \mu \delta_{jl}. \label{eq:D_iso} 
\end{gather}
By comparing the above to \eqref{eq:C_reg_hex}, \eqref{eq:B_reg_hex}, and 
\eqref{eq:D_reg_hex}, we see that the Coble creep model of regular hexagons 
here is a special case of the general two-dimensional linear isotropic 
micropolar continuum with $\zeta = 2 \eta$, and $\kappa = 0$. The degeneracy in 
the rheology is that associated with $\kappa =0$. Substitution of 
\eqref{eq:C_iso}, \eqref{eq:B_iso}, and \eqref{eq:D_iso} into \eqref{eq:con1} 
and \eqref{eq:con2} yields the isotropic constitutive laws
\begin{align}
 \sigma_{ij} &= \zeta \Gamma_{kk} \delta_{ij} + 2 \eta \Gamma_{\lbrace(ij)\rbrace} + 
2 \kappa \Gamma_{[ ij ]}, \label{eq:sig_cons} \\
\chi_{3j} &= \mu K_{3j}, \label{eq:chi_cons}
\end{align}
where $\Gamma_{\lbrace(ij) \rbrace}$ denotes the trace-free part of the symmetric tensor $\Gamma_{(ij)}$, given explicitly in 2D by
\begin{equation}
\Gamma_{\lbrace(ij)\rbrace} = \frac{1}{2} \left(\Gamma_{ij} + \Gamma_{ji} -
\Gamma_{kk} \delta_{ij}  \right).
\end{equation}
$\zeta$ and $\eta$ represent the conventional bulk and shear viscosities; 
$\kappa$ and $\mu$ are the new constants associated with the micropolar medium. 
We will refer to $\kappa$ as the microrotational viscosity, and $\mu$ as the 
angular viscosity. Substitution of the constitutive laws \eqref{eq:sig_cons} 
and \eqref{eq:chi_cons} into the conservation laws \eqref{eq:micropolar_mom} and 
\eqref{eq:micropolar_ang_mom} leads to the 2D isotropic governing equations
\begin{gather}
 \zeta \nabla \left( \nabla \cdot \mathbf{v} \right) + \eta \nabla^2 \mathbf{v} 
+ \kappa \nabla \times \left( 2 \boldsymbol{\omega} - \nabla \times \mathbf{v} 
\right) = \boldsymbol{0}, \label{eq:2d_mom}\\
\mu \nabla^2 \boldsymbol{\omega} -2 \kappa  \left( 2 \boldsymbol{\omega} - 
\nabla \times \mathbf{v}  \right)  = \boldsymbol{0}. \label{eq:2d_ang_mom}
\end{gather}
Note that since in 2D, $\mathbf{v} = (v_1, v_2, 0)$ and $\boldsymbol{\omega} = (0, 0, \omega_3)$, the angular momentum balance in \eqref{eq:2d_ang_mom} reduces to a scalar equation for $\omega_3$. When $\kappa=0$, the governing equation for the velocity field $\mathbf{v}$ in \eqref{eq:2d_mom} is decoupled 
from that for the microrotation rate field $\boldsymbol{\omega}$ in \eqref{eq:2d_ang_mom}. When $\kappa=0$, 
\eqref{eq:sig_cons} shows that the stress tensor is symmetric, and the 
combination of \eqref{eq:sig_cons} and \eqref{eq:micropolar_mom} gives the 
2D Navier equation for the velocity field $\mathbf{v}$. Equations 
\eqref{eq:chi_cons} and \eqref{eq:micropolar_ang_mom} then lead simply to 
a 2D Laplace's equation for the microrotation rate $\omega_3$. 

To formulate a full problem for the flow of a micropolar fluid, the governing equations in \eqref{eq:2d_mom} and \eqref{eq:2d_ang_mom} have to be supplemented with boundary conditions. For a micropolar fluid, possible choices include setting on the boundaries the velocities $\mathbf{v}$ and microrotation-rate $\boldsymbol{\omega}$, or the tractions $\tensor{\sigma} \cdot \mathbf{n}$ and the moments $\tensor{\chi} \cdot \mathbf{n}$. For the 2D case, only the $\omega_3$ component of the microrotation-rate is non-zero, and the corresponding non-zero moment is $\mu\, \partial{\omega_3}/\partial n$. 

Even with $\kappa=0$, \eqref{eq:2d_mom} and \eqref{eq:2d_ang_mom} can be solved to give unique velocity and microrotation fields if the velocities and microrotations are set along the boundary. However, if instead the tractions and moments are specified at the boundaries, then the solutions are non-unique. With $\kappa=0$, the solution for $\mathbf{v}$ in the Navier equation in $\eqref{eq:2d_mom}$ is then only unique up to a rigid body motion, and the solution to Laplace's equation for $\omega_3$ in $\eqref{eq:2d_ang_mom}$ is only unique up to an arbitrary constant, and thus the microrotations can be arbitrarily large. This non-uniqueness under certain choices of boundary condition has been noted in previous grain-scale models of Coble creep \citep[e.g.][]{Hazzledine1993,Weber1996,Ford2002}. 

To see why the microrotation viscosity $\kappa$ vanishes for regular hexagonal grains, consider a situation where all the grain centres are fixed, and all grains have the same constant microrotation rate $\boldsymbol{\omega}$. At the grain boundaries the relative motion between grains is purely tangential, and thus involves no plating, and no dissipation. There is thus no resistance to such a motion in the model. The plating rate is independent of the antisymmetric tensor $\Gamma_{[ ij ]}$.

The independence of the plating rate (and thus the constitutive laws) on the 
antisymmetric tensor $\Gamma_{[ ij ]}$ will be seen for a number of grain 
geometries, in particular those geometries where the vectors $\mathbf{R}$ 
joining grain centres are parallel to the normals $\mathbf{n}$ of the grain 
boundaries. Mathematically, this is straightforward to see from the plating rate 
expression in \eqref{eq:plating}: the second rank tensor $R_i n_j$ is a 
symmetric tensor if $\mathbf{R}$ is parallel to $\mathbf{n}$, and thus will 
yield zero when contracted with the antisymmetric tensor $\Gamma_{[ ij ]}$. 
Again, with the grain centres fixed, a uniform microrotation of the grains leads 
to purely tangential motion at the grain boundary. 

In 3D there is an additional degeneracy in the rheology. When $\mathbf{R}$ and $\mathbf{n}$ are parallel, the plating rate \eqref{eq:plating} is also independent of $K_{kk} = \nabla \cdot \boldsymbol{\omega}$, and thus so is the dissipation. This arises because a relative rotation of neighbouring grains about an axis perpendicular to the grain boundary involves purely tangential motion at the boundary, and thus no plating. Such a mode of deformation does not occur in 2D. For a 2D isotropic medium there is only a single angular viscosity coefficient that describes the resistance to relative rotation of grains; for a 3D isotropic medium there are in general three such viscosity coefficients, where the one associated with $\nabla \cdot \boldsymbol{\omega}$ will vanish when $\mathbf{R}$ and $\mathbf{n}$ are parallel.

A wide range of artificial grain geometries have the property that the vectors $\mathbf{R}$ joining grain centres are parallel to the normals $\mathbf{n}$ of the grain boundaries. For example, an artificial grain geometry could be obtained from a Voronoi tessellation, and this tessellation has the property that the lines joining the generator points across each cell boundary are normal to the boundary, and thus allow pure tangential motion to occur with a uniform microrotation of the grains \citep{Weber1996}. Moreover, generalisations of the Voronoi tessellation, such as Laguerre tessellations, also have this property, and thus grain geometries produced by software such as Neper \citep{Quey2018} would also allow uniform microrotations of grains about their generator points to occur without plating. In 3D, all normal tessellations with convex grains are Laguerre tessellations (see Theorem 3.2 of \citep{Lautensack2008}).  

However, not all artificial grain geometries have $\mathbf{R}$ and $\mathbf{n}$ parallel, and indeed the 2D irregular hexagons described by Wheeler \citep{Wheeler2010} give an example where $\mathbf{R}$ and $\mathbf{n}$ are not parallel. The tiling of irregular hexagons is not in a general a Voronoi or Laguerre tiling, although a subset are. Even for the geometries where $\mathbf{R}$ and $\mathbf{n}$ are not parallel there are modes of deformation that can occur without plating. However, these modes now include both the symmetric and antisymmetric parts of the tensor $\Gamma_{ij}$. It was shown in \citep{Wheeler2010} that for the irregular hexagon any constant multiple of 
\begin{equation}
 \tensor{\Gamma} = \left(\widehat{\mathbf{B}}_2 \cdot \mathbf{N}_3 \right) \mathbf{B_1} \otimes \mathbf{N}_1 + \left(\widehat{\mathbf{B}}_3 \cdot \mathbf{N}_1 \right) \mathbf{B_2} \otimes \mathbf{N}_2 + \left(\widehat{\mathbf{B}}_1 \cdot \mathbf{N}_2 \right) \mathbf{B_3} \otimes \mathbf{N}_3
\end{equation}
can occur without plating. 

\subsection{Grain boundary sliding}

It seems rather unsatisfactory to have a rheological model that allows some modes of deformation to occur without any resistance. It suggests that additional physical processes should be considered that describe resistance to such modes. An obvious example of such an additional process is the resistance to grain boundary sliding \citep{Lifshitz1963,Beere1978}. This is often modelled as if there were a thin layer of fluid of viscosity $\eta_{\text{gb}}$ at the grain boundary that resists shearing motion. The effective viscosity tensors for this process can be derived in exactly the same way as those for the plating resistance by matching the dissipation, and the detailed calculations are given in \autoref{sec:gb}. In the case of 2D regular hexagons, the relevant viscosities including both plating resistance and sliding resistance are given in dimensional form as 
\begin{align}
 \zeta &= \frac{1}{72} \frac{k T d^3}{\delta D^\text{gb} \Omega},  \label{eq:hex_bulk} \\
 \eta &= \frac{1}{144} \frac{k T d^3}{\delta D^\text{gb} \Omega} + \frac{1}{4} \eta_\text{gb} \frac{ d}{\delta},  \label{eq:hex_shear} \\
 \kappa &= \frac{1}{2} \eta_\text{gb} \frac{ d}{\delta}, \label{eq:hex_rot} \\
 \mu &= \frac{1}{6480} \frac{k T d^5}{\delta D^\text{gb} \Omega} + \frac{1}{4} \eta_\text{gb} \frac{ d^3}{\delta}. \label{eq:hex_ang}
\end{align}

\begin{figure}
 \includegraphics[width=\columnwidth]{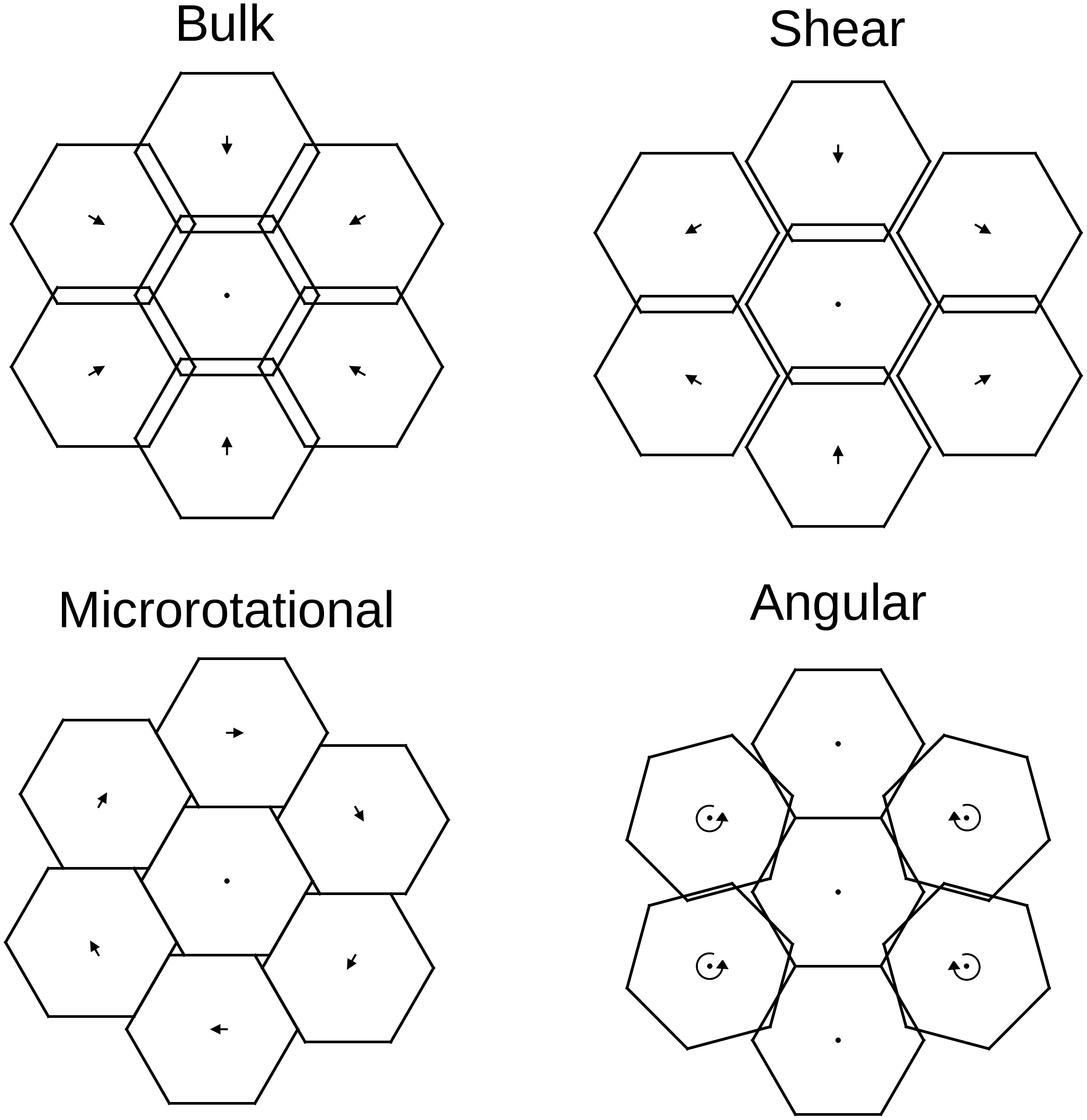}
 \caption{Examples of the different modes of deformation for a tiling of regular hexagons. In each case the frame of reference is such that the central grain is stationary (neither translating nor rotating). Straight arrows indicate the velocities of the centres of each grain, curved arrows indicate the sense of grain rotation. The hexagonal grains have been displaced by finite amounts to show their motions, but the theory described in the text concerns infinitesimal motions only. Regions of overlap and gaps between grains must be accommodated by plating out or removal of material. Bulk deformation involves purely plating; there is no sliding between grains. Shear deformation involves both plating and sliding. Microrotational deformation involves pure sliding; in the example shown none of the individual grains rotate, but the grain centres undergo a rotation. Angular deformation occurs when there is relative rotation between neighbouring grains; such deformation involves both plating and sliding.}
 \label{fig:def_modes}
\end{figure}
\autoref{fig:def_modes} illustrates the grain-scale deformation associated with each viscosity coefficient. The expressions for the effective bulk and shear viscosity for regular hexagons are well-established results (e.g. \citep{Spingarn1978, Beere1978, Cocks1990}); the novel results here are explicitly identifying the microrotational viscosity $\kappa$ and the angular viscosity $\mu$. Both the plating and sliding resistance to relative rotation of grains has been described by other authors in the context of determining the rotation rate of an individual grain when a torque is applied  \citep{Moldovan2001,Harris1998,Kim2005}. The novelty here is to relate this resistance to relative rotation to the angular viscosity $\mu$ of a micropolar fluid. The inclusion of resistance to grain boundary sliding leads to a non-zero microrotational viscosity $\kappa$, and formally removes the degeneracy in the rheology. However, having significant resistance to grain boundary sliding also changes the effective shear viscosity \eqref{eq:hex_shear}, with the consequence that if $ \eta_\text{gb} \gg k T d^2 / (D^\text{gb} \Omega)$ the effective shear viscosity depends linearly on the grain size rather than the cubic dependence typical for Coble creep. But if the resistance to sliding is weak, one will still then find modes of deformation which are only weakly resisted \citep{Wheeler2010}. 

An important feature of a micropolar continuum that is distinct from a classical Cauchy continuum is the existence of a characteristic length scale of the medium. One way of defining this characteristic length scale is from the balance of the terms $\mu \nabla^2 \boldsymbol{\omega}$ and  $4 \kappa \boldsymbol{\omega}$ in equation \eqref{eq:2d_ang_mom}, which defines the characteristic length scale
\begin{equation}
 l = \frac{1}{2} \sqrt{\frac{\mu}{\kappa}}. \label{eq:char_length}
\end{equation}
The behaviour of the micropolar medium is different depending on whether the imposed scales of deformation $L$ are significantly larger or smaller than the characteristic length scale $l$ of the medium. If $L \gg l$, then the  resistance to relative rotation of the grains can be neglected throughout most of the fluid, with the exception of thin boundary layers of thickness $l$. Significant gradients in the microrotation rate only occur across these boundary layers. Outside these boundary layers, the equations describing the medium can simplified to those of a reduced micropolar continuum (see \autoref{sec:reduced} for a detailed discussion of this limit). On the other hand, if $L \ll l$, then angular resistance dominates, and a gradient in microrotation rate can exist across the domain.  

The characteristic length scale $l$ depends on the grain size, but is potentially much larger than the grain size. Consider the case where resistance to grain-boundary sliding is weak, $\eta_\text{gb} \ll k T d^2 / (D^\text{gb} \Omega)$. Substitution of \eqref{eq:hex_rot} and \eqref{eq:hex_ang} into \eqref{eq:char_length} then yields
\begin{equation}
 l = \frac{d^2}{36 \sqrt{10}} \sqrt{\frac{k T}{\eta_\text{gb} D^\text{gb} \Omega}} \label{eq:char_length2}
\end{equation}
which demonstrates explicitly that the characteristic length scale of the micropolar medium is not simply the grain size. Another way of writing \eqref{eq:char_length2} is in terms of the effective shear viscosity for pure Coble creep, $\eta_\text{Coble} = k T d^3/(144 \delta D^\text{gb} \Omega)$, and for pure grain-boundary-sliding, $\eta_\text{gbs} = \eta_\text{gb} d/(4 \delta)$,
\begin{equation}
 l = \frac{d}{6 \sqrt{10}} \sqrt{\frac{\eta_\text{Coble}}{\eta_\text{gbs}}}
\end{equation}
which demonstrates that the characteristic length scale $l$ can be significantly greater than the grain size if the resistance to grain boundary sliding is weak ($\eta_\text{gbs} \ll \eta_\text{Coble}$).

\subsection{Constraints}

An alternative method for removing the degeneracy without invoking resistance to grain boundary sliding is by imposing additional constraints on the allowable deformation. A simple example of a constraint is that of incompressibility where one demands that $\nabla \cdot \mathbf{v} = 0$ everywhere, or equivalently that $\Gamma_{kk} = 0$, and the bulk viscosity $\zeta$ becomes effectively infinite (see appendix \ref{sec:constraints}). A similar constraint is added in some micropolar models to demand that rotations are constrained as $\boldsymbol{\omega} = \tfrac{1}{2} \nabla \times \mathbf{v}$ everywhere, or equivalently that $\Gamma_{[ij]} = 0$, and the microrotational viscosity $\kappa$ becomes effectively infinite. 

In adding constraints on the deformation, one should justify them based on the microscale physics. It is straightforward to consider microphysical assumptions that impose incompressibility. While here the constitutive laws were derived under the assumption that the triple lines where grain boundary meet act as sources and sinks of vacancies \citep{Cocks1990,Takei2009,Rudge2018a}, one can instead consider microphysical models where this is not the case and one has to balance fluxes of vacancies at the triple lines and this naturally leads to a incompressible rheology \citep{Spingarn1978, Wheeler2010, Rudge2018a}. While having constrained rotations would remove the degeneracy without invoking the resistance to grain boundary sliding, there does not seem to be an obvious micromechanical mechanism that imposes such a constraint.

\section{Conclusions}\label{sec:conclusions}

Here we have presented a new perspective on diffusion creep by identifying an equivalent micropolar fluid model that shares the same dissipation and kinematics as the discrete collection of grains. We have shown how the effective constitutive tensors can be obtained for both Nabarro-Herring (volume diffusion) and Coble (grain-boundary diffusion) creep. Specific calculations have been performed for a periodic array of hexagons undergoing Coble creep in two dimensions. 

For most practical purposes identifying a material undergoing diffusion creep with a Newtonian viscous fluid in the classical Cauchy continuum framework is appropriate. However, there are cases where using the more general micropolar framework developed here may be beneficial. Most obviously, the addition of rotational degrees of freedom to the continuum description allows one to describe grain rotation. Including  micropolar effects will be particularly important whenever the length scales of deformation approach those of the characteristic length scale of the micropolar medium; this may be important in problems where localisation of deformation occurs (e.g. shear banding). Even in problems where deformation occurs at length scales much larger than the characteristic length scale, inclusion of rotational degree of freedoms are important for determining the effective viscosities when the microstructure is anisotropic (Appendix \ref{sec:reduced}, \citep{Trovalusci2003,Fantuzzi2020}).       

There are several natural avenues for future research. In this work specific calculations were only performed in the very simplest case of Coble creep in two-dimensions because it is straightforward to solve the relevant equations analytically. Detailed calculations could be performed in 3D for both Coble creep and Nabarro-Herring creep, but these involve numerical computation. We demonstrated that consideration of plating resistance alone leads to a degenerate rheology, and consideration of other physical processes, such as the resistance to grain boundary sliding, are necessary to remove this degeneracy. The physics of grain boundaries were treated here in a very simple-minded way,  offering just a Newtonian viscous resistance to sliding. Future work should consider a better description of the physics of the grain boundaries, e.g. a detailed consideration of grain boundary energies and grain boundary migration. We have also addressed here only the instantaneous deformation of the material; there is a lot of work that  still that needs to be done to explore how the microstructure evolves during deformation, and how that influences subsequent deformation. Finally, there is the prospect that future laboratory experiments may place constraints on the appropriate micropolar constitutive tensors for diffusion creep.

\bibliographystyle{tfq}

\appendix

\section{Resistance to grain-boundary sliding}\label{sec:gb}

A simple way to model grain-boundary sliding is to treat the interface between 
grains as if it were a thin layer of viscous fluid that offers resistance to 
being sheared \citep[e.g.][]{Lifshitz1963, Beere1978, Kim2009}. The resistance can be characterised by 
averaged dissipation over the grain in the form
\begin{equation}
 \Psi = \frac{\eta_\text{gb}}{2 V \delta} \int | \mathbf{n} \times \Delta \v 
|^2 \, \d S \label{eq:shear_diss}
\end{equation}
where $\eta_\text{gb}$ is the effective shear viscosity of the grain boundary, 
and $\delta$ is its effective thickness. Using the notion of equivalent 
dissipation, the effective constitutive tensors can be obtained by matching 
\eqref{eq:shear_diss} and \eqref{eq:deltav3} with \eqref{eq:mpolar_diss}. In what follows we will use 
dimensionless variables, where $C_{ijkl}$ is scaled on $\eta_\text{gb} d / 
\delta$, $B_{ijkl}$ on $\eta_\text{gb} d^2 / 
\delta$ and $D_{ijkl}$ on $\eta_\text{gb} d^3 / 
\delta$. From \eqref{eq:deltav3} it follows that 
\begin{equation}
 \left(\mathbf{n} \times \Delta \v \right)_p =a_{pkl} \Gamma_{kl}  + b_{pkl} K_{kl} 
\end{equation}
where
\begin{gather}
 a_{pkl} = \epsilon_{p q k} n_q R_l, \\
 b_{pkl} = c_{pk} R_l, \\
 c_{pk} = \left( \dv \cdot \n \right) \delta_{pk} - d_p n_k.
\end{gather}
It follows that
\begin{gather}
 C_{ijkl} = \frac{1}{2 V} \int \left( \delta_{ik} - n_i n_k \right) R_j 
R_l 
\, \d S, \label{eq:C_b} \\
B_{ijkl} = - \frac{1}{2V} \int \left(\mathbf{d} \times \mathbf{n} \right)_i n_k R_j R_l \, \d S, \label{eq:B_b} \\
D_{ijkl} = \frac{1}{2V} \int c_{pi} c_{pk} R_j R_l \, \d S. \label{eq:D_b}
\end{gather}

\subsection{Irregular hexagons}

For the irregular hexagons described in \autoref{sec:hex}, the expressions in \eqref{eq:C_b}, \eqref{eq:B_b}, and \eqref{eq:D_b} reduce to  
\begin{gather}
 C_{ijkl} = \frac{1}{V} \sum_{f=1}^3 \left(\delta_{ik} - N^f_i N^f_k
\right) R^f_j R^f_l l_f,  \\
B_{3jkl} = \frac{1}{2 V} \sum_{f=1}^3 \left(\mathbf{R}_f \cdot \mathbf{B}_f \right)  N^f_k 
R^f_j R^f_l, \\
D_{3j3l} = \frac{1}{4 V} \sum_{f=1}^3  \left(\mathbf{R}_f \cdot  \mathbf{N}_f \right)^2   R^f_j R^f_l l_f.
\end{gather}

\subsection{Regular hexagons}

In the special case of regular hexagons, isotropy imposes that the constitutive tensors take the form in \eqref{eq:C_iso}, \eqref{eq:B_iso}, and \eqref{eq:D_iso} in terms of the four constants $\zeta$, $\eta$, $\kappa$, and $\mu$. These constants can be determined from the tensor invariants,
\begin{gather}
 C_{iikk} = 4 \zeta = 0, \\
 C_{ijij} = 2 \zeta + 4 \eta + 2 \kappa = 2, \\
 C_{ijji} = 2 \zeta + 4 \eta - 2 \kappa = 0, \\
 D_{3j3j} = 2 \mu = \tfrac{1}{2},
\end{gather}
where the units are such that the perpendicular distance between opposite sides of the hexagon equals 1. It follows that
\begin{equation}
\zeta = 0, \quad  \eta = \frac{1}{4},\quad \kappa = \frac{1}{2},\quad \mu = \frac{1}{4},
\end{equation}
where it should be noted that $\zeta$, $\eta$, and $\kappa$ are scaled by $\eta_\text{gb} d / 
\delta$ and $\mu$ by $\eta_\text{gb} d^3 / 
\delta$. The result that $\eta = 1/4$ is in complete agreement with the results obtained by \citep{Beere1978,Kim2009,Wheeler2010}. The new results here are for the micropolar constants $\kappa$ and $\mu$. 

\section{Reduced micropolar continuum}\label{sec:reduced}

A micropolar medium has an intrinsic characteristic length scale $l$. If variations in the flow happen on much longer scales, say a length scale $L$, with $L \gg l$, then the constitutive laws can be further simplified to a reduced micropolar model outside any boundary layers (e.g. \citep{Grekova2009,Grekova2019}). If $\eta$ is a typical viscosity, the constitutive tensors in \eqref{eq:con1} and \eqref{eq:con2} scale as $C_{ijkl} \sim \eta$, $B_{ijkl} \sim \eta l$, and $D_{ijkl} \sim \eta l^2$. If $\dot{\varepsilon}$ is a typical strain rate, then $\Gamma_{ij} \sim \dot{\varepsilon}$ and $K_{ij} \sim \dot{\varepsilon}/L$. By keeping terms only at leading order in the small parameter $l/L$, the balance laws \eqref{eq:micropolar_mom} and \eqref{eq:micropolar_ang_mom} become
\begin{gather}
 \sigma_{ij,j}  = 0,  \label{eq:reduced_micropolar_mom} \\
  \epsilon_{ijk} \sigma_{jk}  = 0, \label{eq:reduced_micropolar_ang_mom}
\end{gather}
and the constitutive law simplifies to 
\begin{equation}
 \sigma_{ij} = C_{ijkl} \Gamma_{kl}.  \label{eq:reduced_con} 
\end{equation}
In this reduced micropolar continuum, couple stresses are negligible ($\chi_{ij} = 0$), the dissipation is independent of $K_{ij}$ (the gradient in microrotation rate), and the force-stress tensor $\sigma_{ij}$ is symmetric (equation \eqref{eq:reduced_micropolar_ang_mom}). It was under these assumptions that the calculations of Wheeler \citep{Wheeler2010} were made, and we will show now that the more general micropolar continuum considered here reduces to that of Wheeler \citep{Wheeler2010} with these additional assumptions.

It is helpful to decompose both the Cosserat tensor $\Gamma_{ij}$ and the force-stress tensor $\sigma_{ij}$ into their symmetric and antisymmetric parts, as illustrated for $\Gamma_{ij}$ in equations \eqref{eq:gam_as1}, \eqref{eq:gam_as2}, and \eqref{eq:rel_micro}. The constitutive law in \eqref{eq:reduced_con} can be written as 
\begin{gather}
 \sigma_{(ij)} = C_{(ij)(kl)} \Gamma_{(kl)} + C_{(ij)[ kl ]} \Gamma_{[ kl ]},  \label{eq:redcon1} \\
 \sigma_{[ ij ]} = C_{[ ij ](kl)} \Gamma_{(kl)} + C_{[ ij ] [ kl ]} \Gamma_{[ kl ]}, \label{eq:redcon2}
\end{gather}
where round brackets $(ij)$ represent the symmetric part, and square brackets $[ ij]$ represent the antisymmetric part. Note that due to the major symmetry $C_{ijkl}=C_{klij}$, we have that $C_{[ ij ](kl)} = C_{(kl) [ ij ]}$.

Antisymmetric second-rank tensors can be conveniently represented using pseudo-vectors, which we will define for the Cosserat tensor $\Gamma_{ij}$ as
\begin{gather}
 \domega_k = \frac{1}{2} \epsilon_{kij} \Gamma_{ij}, \\
 \Gamma_{[ ij]} = \epsilon_{ijk} \domega_k, \label{eq:rel_micro2}
\end{gather}
using the superscript $\times$ to denote the pseudo-vector. The pseudo-vector $\dsigma_k$ similarly relates to the antisymmetric part of the force-stress tensor $\sigma_{[ij]}$. Note from \eqref{eq:rel_micro} that $ \domega_k = \omega_k - \tfrac{1}{2} 
\Omega_k$, and so $\domega_k$ represents the relative rate of grain rotation with respect to the rotation of the grain centres.

The viscosity tensors in \eqref{eq:redcon1} and \eqref{eq:redcon2} can be simplified by introducing a symmetric second-rank tensor $A_{pq}$ as
\begin{gather}
 A_{pq} = \frac{1}{4} \epsilon_{pij} \epsilon_{qkl} C_{ijkl},\\
C_{[ ij] [ kl ]} = \epsilon_{pij} \epsilon_{qkl} A_{pq}, 
 \end{gather}
and a third-rank pseudo-tensor $E_{pij}$ as
\begin{gather}
 E_{pij} = \frac{1}{2} \epsilon_{pkl} C_{(ij)kl}, \\
 C_{(ij)[ k l ]} = \epsilon_{pkl} E_{pij}.
\end{gather}
Note that $E_{pij}$ is symmetric on its last two indices, $E_{pij} = E_{pji}$. For an isotropic medium, $A_{pq} = \kappa \delta_{pq}$ where $\kappa$ is the microrotation viscosity, and $E_{pij}$ vanishes. The constitutive laws in \eqref{eq:redcon1} and \eqref{eq:redcon2} can be written as
\begin{gather}
 \sigma_{(ij)} = C_{(ij)(kl)} \Gamma_{(kl)} + 2 E_{kij} \domega_k, \label{eq:redcon3}\\
\dsigma_i = E_{ikl} \Gamma_{(kl)} + 2 A_{ik} \domega_k. \label{eq:redcon4}
\end{gather}
Equation \eqref{eq:reduced_micropolar_ang_mom} implies that $\dsigma_i=0$, which when combined with \eqref{eq:redcon4} places a constraint on the allowable motions. If the second-rank symmetric tensor $A_{ik}$ is invertible, this constraint can be explicitly written as giving the relative rotation rate pseudo-vector $\domega_k$ in terms of the symmetric strain-rate tensor as
\begin{equation}
 \domega_k = - \frac{1}{2} A^{-1}_{kp} E_{pij} \Gamma_{(ij)}. \label{eq:rotrate}
\end{equation}
The result in \eqref{eq:rotrate} is analogous to the expressions which determine the rotation rate of a rigid particle embedded in a viscous fluid undergoing shear (e.g. \citep{Graham2018,Guazzelli2011}). In these analogous problems the torque on the rigid particle can be related to the relative rotation rate of the particle and the far-field strain rate through resistance tensors, and the particle rotation rate can be determined by a condition of zero net torque. Equation \eqref{eq:rotrate} can substituted into the constitutive law \eqref{eq:redcon3} for the symmetric part of the stress tensor and thus show that the medium behaves as a classical Cauchy continuum governed by
\begin{gather}
 \sigma_{(ij), j}  = 0, \\
 \sigma_{(ij)} = \tilde{C}_{ijkl} \Gamma_{(kl)}, \\
 \tilde{C}_{ijkl} = C_{(ij)(kl)} - E_{pij} A^{-1}_{pq} E_{qkl}, \label{eq:wheelercon}
\end{gather}
where the degrees of freedom associated with the microrotations have been eliminated. Note that the effective viscosity tensor $\tilde{C}_{ijkl}$ satisfies both the major and the minor symmetries ($\tilde{C}_{ijkl}=\tilde{C}_{jikl}$), and that it differs in general from $C_{(ij)(kl)}$. Thus while the effective medium behaves as a Cauchy continuum, one cannot neglect the rotational degrees of freedom when determining the effective viscosity tensor. Only when the third-rank pseudo-tensor $E_{kij}$ vanishes will $\tilde{C}_{ijkl}=C_{(ij)(kl)}$, which can occur when certain symmetries are present. For example, in 3D, $E_{kij}$ vanishes when the grain geometry is invariant under the crystallographic point groups $432$, $\overline{4}3\textit{m}$, or \textit{m}$\overline{3}$\textit{m}. If only plating resistance is considered, both $E_{kij}$ and $A_{ij}$ vanish for any grain geometry for which the vector $\mathbf{R}$ joining grain centres is parallel to the grain boundary normal $\mathbf{n}$, as the plating rate (and thus dissipation) is then independent of $\domega_k$. When $E_{kij}$ vanishes, the translational motion is decoupled from the rotational motion, and provided $A_{ij}$ is invertible, the microrotation rate is constrained to be half the vorticity, $\boldsymbol{\omega} = \tfrac{1}{2} \left(\nabla \times \mathbf{v} \right)$. If, however, both the tensor $E_{kij}$ and the 
tensor $A_{ij}$ vanish (e.g. zero microrotation viscosity $\kappa$ in the isotropic case), the microrotation rate $\boldsymbol{\omega}$ is unconstrained.

When the grain geometry has $\mathbf{R}$  parallel to $\mathbf{n}$, and the resistance to grain boundary sliding is included, but relatively weak compared with the plating resistance, then the effective viscosity tensor is dominated by the plating resistance, with $\tilde{C}_{ijkl} \approx C_{(ij)(kl)}$. However, the relative rotation rate $\domega_k$ of the grains is determined purely by the resistance to grain boundary sliding (\autoref{sec:gb}), as the tensors $A_{ij}$ and $E_{kij}$ depend only on the sliding resistance. The third-rank pseudo-tensor $A^{-1}_{kp} E_{pij}$ in \eqref{eq:rotrate} is independent of the grain boundary sliding viscosity $\eta_\text{gb}$ and depends only on the geometry of the grain.

Wheeler \citep{Wheeler2010} performed the same eliminations leading to \eqref{eq:rotrate} and \eqref{eq:wheelercon} for the specific case of hexagons in 2D. In a 2D medium, there is only a single non-zero component $\Gamma_3^\times$ representing the relative rotation rate, and 
only the $A_{33}$ and $E_{3ij}$ components of the appropriate viscosity tensors are non-zero. The 2D eliminations yield
\begin{gather}
\domega_3 = - \frac{1}{2} \frac{E_{3ij}}{A_{33}} \Gamma_{(ij)}, \label{eq:rotrate2d} \\
  \tilde{C}_{ijkl} = C_{(ij)(kl)} - \frac{E_{3ij} E_{3kl}}{A_{33}}, \label{eq:wheelercon2d}
\end{gather}
and the dimensionless tensors describing plating resistance for hexagons are given explicitly as
\begin{gather}
C_{(ij)(kl)} = \frac{1}{48 V} \sum_{f=1}^3 \left(N^f_i R^f_j + N^f_j R^f_i \right) \left(N^f_k R^f_l + N^f_l R^f_k \right)    l_f^3, 
\label{eq:Csym_hex}\\ 
E_{3ij} = \frac{1}{24 V} \sum_{f=1}^3 \left(N^f_i R^f_j + N^f_j R^f_i \right) 
\left(\mathbf{N}_f \times \mathbf{R}_f\right)_3 l_f^3, \label{eq:Ehex}\\
A_{33} = \frac{1}{48 V} \sum_{f=1}^3 \left(\mathbf{N}_f \times \mathbf{R}_f\right)_3  
\left(\mathbf{N}_f \times \mathbf{R}_f\right)_3 l_f^3.\label{eq:A_hex} 
\end{gather}
Equation (23) in \citep{Wheeler2010} is a special case of \eqref{eq:rotrate2d}, and equation \eqref{eq:wheelercon2d} can be used to reproduce the same constitutive laws as found by \citep{Wheeler2010} for the cases where $A_{33} \neq 0$. As remarked above, for plating resistance both $E_{3ij}$ and $A_{33}$ vanish when $\mathbf{N}_f$ and $\mathbf{R}_f$ are parallel, which can be seen in the expressions in \eqref{eq:Ehex} and \eqref{eq:A_hex} in the dependence on the cross product $\mathbf{N}_f \times \mathbf{R}_f$. In such cases, one cannot make the elimination leading to \eqref{eq:wheelercon2d}, and instead we simply have that the effective viscosity tensor is $\tilde{C}_{ijkl} = C_{(ij)(kl)}$. Wheeler \citep{Wheeler2010} states that ``no unique value of
viscosity exists for the rheology of an array of regular hexagons'' when only plating resistance is considered. He arrives at this conclusion based on taking different limits of an elimination similar to that in \eqref{eq:wheelercon2d}; however such an elimination is not possible when $A_{33}$ vanishes. By Neumann's principle the effective viscosity tensor of an array of regular hexagonal grains is always isotropic. The shear viscosity is well defined and strictly positive; the degeneracy is simply that the microrotational viscosity vanishes when the only resistance to motion is plating.

\section{Constraints}\label{sec:constraints}

\subsection{Incompressibility}

Material models can be supplemented by constraints which forbid certain modes of 
deformation. An important example of this is an incompressibility constraint, 
which forbids bulk deformation ($\nabla \cdot \mathbf{v} = \Gamma_{kk} = 0$). 
The viscosity tensors here are derived assuming bulk deformation is possible, 
with potential sources and sinks of vacancies where grain boundaries meet 
\citep{Cocks1990,Rudge2018a,Takei2009}. The bulk deformation can be separated 
from the rest of the deformation by splitting the Cosserat tensor $\Gamma_{ij}$ 
into an isotropic part and a trace-free part as 
\begin{gather}
 \Gamma_{ij} = \frac{1}{N}\Gamma_{kk}\delta_{ij} + \Gamma_{\lbrace ij \rbrace} \\
 \Gamma_{\lbrace ij \rbrace} = \Gamma_{ij} - \frac{1}{N} \Gamma_{kk} \delta_{ij},
\end{gather}
where $N$ is the dimension of the space. Here curly brackets indicate a tensor that yields zero when contracted over the indices within the curly brackets.

For an incompressible medium, the dissipation depends only on the trace-free tensor $\Gamma_{\lbrace ij \rbrace}$.  The constitutive laws in \eqref{eq:con1} and \eqref{eq:con2} are replaced by   
\begin{gather}
\sigma_{ij} = -P \delta_{ij} + \sigma_{\lbrace ij \rbrace}, \\
 \sigma_{\lbrace ij \rbrace } =  C_{\lbrace ij \rbrace \lbrace kl \rbrace} \Gamma_{\lbrace kl \rbrace} + B_{\lbrace ij \rbrace kl} K_{kl}, \label{eq:tfree_con1} \\
  \chi_{ij} = B_{\lbrace kl \rbrace ij} \Gamma_{\lbrace kl \rbrace} + D_{ijkl} K_{kl}, \label{eq:tfree_con2}
\end{gather}
where $P$ is the pressure, a Lagrange multiplier  enforcing the incompressibility constraint \citep{Eremeyev2013}. The viscosity tensors derived here can be reduced to those for an incompressible medium as
\begin{gather}
 C_{\lbrace ij \rbrace \lbrace kl \rbrace} = C_{ijkl} - \frac{1}{N} C_{pp k l} \delta_{ij} -  \frac{1}{N} C_{ij qq} \delta_{kl} + \frac{1}{N^2} C_{pp qq} \delta_{ij}\delta_{kl}, \\
  B_{\lbrace ij \rbrace kl} = B_{ijkl} - \frac{1}{N} B_{pp k l} \delta_{ij}.
\end{gather}
For the 2D isotropic medium considered in \eqref{eq:2d_mom} and \eqref{eq:2d_ang_mom}, the effect of adding an incompressibility constraint is to replace the $\zeta \nabla \left( \nabla \cdot \mathbf{v} \right)$ term in \eqref{eq:2d_mom} by a $-\nabla P$ term. 

\subsection{Constrained rotations}

Another constraint that can be considered is that of constrained rotations, where we demand that $\boldsymbol{\omega} = \tfrac{1}{2} \nabla \times \mathbf{v}$ everywhere, or equivalently that $\Gamma_{[ij]}=0$. It follows that $K_{kk} = \nabla \cdot \boldsymbol{\omega} = 0$. A pseudo-vector Lagrange multiplier $Q_k$ is introduced to enforce the pseudo-vector constraint that $\domega_k = 0$ \citep{Eremeyev2013}. The constitutive laws in \eqref{eq:con1} and \eqref{eq:con2} are replaced by  
\begin{gather}
\sigma_{ij} = Q_k \epsilon_{kij} + \sigma_{(ij )}, \\
 \sigma_{(ij) } =  C_{(ij)(kl)} \Gamma_{(kl)} + B_{(ij) kl} K_{kl}, \label{eq:rot_con1} \\
  \chi_{ij} = B_{(kl)ij} \Gamma_{(kl)} + D_{ijkl} K_{kl}, \label{eq:rot_con2}
\end{gather}
where the Lagrange multiplier represents the antisymmetric part of the stress tensor since $Q_k = \sigma_{k}^{\times}$. For micropolar media with constrained rotations, the microrotation velocity $\boldsymbol{\omega}$ can be eliminated from the governing equations as follows. The antisymmetric part of the force-stress tensor can be eliminated from the conservation laws \eqref{eq:micropolar_mom} and \eqref{eq:micropolar_ang_mom} to yield
\begin{equation}
  \sigma_{(ij),j} + \frac{1}{2}\epsilon_{ijk} \chi_{kl,lj} = 0.
\end{equation}
The above equation, combined with the constitutive laws \eqref{eq:rot_con1} and 
\eqref{eq:rot_con2} describes the micropolar medium with constrained rotations. 
In the case of the 2D isotropic medium considered in \eqref{eq:2d_mom} and 
\eqref{eq:2d_ang_mom} the above can be simplified to
\begin{equation}
 \zeta \nabla \left( \nabla \cdot \mathbf{v} \right) + \eta \nabla^2 \mathbf{v} 
-\frac{\mu}{4} \nabla^4 \mathbf{v} = \boldsymbol{0}. \label{eq:2d_rot_mom}
\end{equation}
This differs from the usual 2D Navier equation by the fourth order term, representing the resistance to relative rotations of the grains. 

\section*{Acknowledgments}

I thank Prof. Samuel Forest for his editorial handling, and two anonymous 
reviewers for their helpful comments.

\end{document}